\documentclass[11pt,a4]{article}
\usepackage{amsmath}
\usepackage{amssymb}
\usepackage{theorem}
\usepackage{color}
\newcommand{\R}{{\cal R}}
\newcommand{\B}{{\cal B}}

\newcommand{\Hc}{{\cal H}}
\newcommand{\Nb}{{\mathbb{N}}}

\newcommand{\qed}{\hfill\hbox{\rule{6pt}{6pt}}}
\newtheorem{theorem}{Theorem}[section]

{\theorembodyfont{\rmfamily}
{\theorembodyfont{\rmfamily}

\begin{document}
\title{Universal test for Hippocratic randomness}
\author{Hayato Takahashi\thanks{1-1 Yanagido, Gifu City 501-1193, Japan. Organization for Promotion of Higher Education and Student Support, Gifu University.  hayato.takahashi@ieee.org}}
\date{\today}
\maketitle

\begin{abstract}
Hippocratic randomness is defined in a similar way to Martin-L\"{o}f randomness, however it does not assume 
computability of the probability and the existence of universal test is not assured.
We introduce the notion of approximation of probability and show  the existence of the universal test (Levin-Schnorr theorem) for Hippocratic randomness when the logarithm of the probability is approximated within additive constant. 
\end{abstract}
\section{Introduction}
In  \cite{{Laurent-etal},{kjoshanssen}}, the notion of Hippocratic (blind) randomness is introduced.
Hippocratic randomness is defined in a similar way to Martin-L\"{o}f randomness, however it does not assume 
computability of the probability. On the other hand if we do not assume computability of the probability, the existence of
universal test is not assured. In this paper, we introduce the notion of approximation of probability and show  the existence of the universal test (Levin-Schnorr theorem) for Hippocratic randomness when the logarithm of the probability is approximated within additive constant.

Let \(\Omega\) be the set of infinite binary sequences and  \(S\) be the set of finite binary strings, respectively. 
Let  \(\Delta(s):=\{sx^\infty \vert x^\infty\in\Omega\}\) for \(s\in S\), where \(sx^\infty\) is the concatenation of \(s\) and \(x^\infty\).
In the following, we study probabilities on \((\Omega, \B)\), where \(\B\) is the  Borel-\(\sigma\)-algebra generated from \(\Delta(s), s\in S\), and
we omit \(\B\) if it is obvious from the context. 
For a probability \(P\) on \(\Omega\), we write \(P(s):=P(\Delta(s))\) for \(s\in S\).

For \(A\subseteq S\), let \(\tilde{A}:=\cup_{s\in A}\Delta(s)\). 
An r.e.~set \(U\subseteq\Nb\times S\) is called  test w.r.t.~\(P\) if \(U_n\supseteq U_{n+1}\) and \(P(\tilde{U}_n)<2^{-n}\), where \(U_n:=\{ x\mid (n,x)\in U\}\), for all \(n\).
The set of Hippocratic random sequences w.r.t.~\(P\) (in the following we denote it by \(\Hc^P\)) is the set that is not covered by any limit of  test, i.e.,  \(\Hc^P:=(\cup_{U: \text{blind test}}\cap_n \tilde{U}_n)^c\).
In this definition, we can replace  \(P(\tilde{U}_n)<2^{-n}\) with  \(P(\tilde{U}_n)<f(n)\), where \(f\) is a computable decreasing function. 
If \(P\) is computable, \(\Hc^P\) is equivalent to Martin-L\"{o}f (ML-)randomness (in the following we denote it by \(\R^P\)).

Next we introduce a notion of approximation.
We say that a probability \(P\) on \(\Omega\) is \(log\)-{\it approximative} if there is a computable \(f:S\to\Nb\) such that
\begin{equation}\label{log-approx}
\exists c\forall x\ f(x)<-\log P(x)<f(x)+c.
\end{equation}
Throughout the paper, the base of logarithm is 2.
For more details of the notion of approximation, see \cite{takahashi2014}.
We can construct an example of  probability on \(\Omega\) that is \(\log\)-approximative but not computable  in a similar manner  to the example (Theorem 2.4)  in \cite{takahashi2014}.

In the following, \(A\subset S\) is called prefix-free (non-overlapping) if \(\Delta(x)\cap\Delta(y)=\emptyset\) for all  \(x,y\in A, x\ne y\).
We see that if \(A\) is r.e.~then there is a prefix-free r.e.~set \(A'\) such that \(\tilde{A}=\tilde{A}'\).
Let \(K_m\) be the 1-dimensional monotone complexity, see \cite{{levin73},{LV2008},{USS90}}.
For a prefix-free set \(A\) we have \(\sum_{x\in A}2^{-Km(x)}\leq 1\), see \cite{takahashiIandC2}.
The following theorem shows that Levin-Schnorr theorem\cite{{levin73},{schnorr73},{schnorr77}} holds for Hippocratic randomness if the probability is \(\log\)-approximative. 

\begin{theorem}[Levin-Schnorr theorem for Hippocratic randomness]
Let \(P\) be a \(\log\)-approximative probability on \(\Omega\). Then
\[x^\infty\in\Hc^P\Leftrightarrow \sup_{x\sqsubset x^\infty} -\log P(x)-Km(x)<\infty.\]
\end{theorem}
Proof)
Suppose that \(x^\infty\notin\Hc^P\).
Then there is a test \(U\) such that \(x^\infty\in\tilde{U}_n\) and \(P(\tilde{U}_n)<2^{-n}\) for all \(n\).
Then there is an r.e.~set \(U'\) such that \(\tilde{U}_n=\tilde{U}'_n\) and \(U'_n\) is prefix-free for all \(n\), where \(U_n^{\prime}:=\{x\mid (n,x)\in U'\}\).
Let \(P'(x)=P(x)2^{n}\) for \(x\in U^{\prime}_n\) and \(0\) otherwise. From (\ref{log-approx}), we have
\[\sum_{x\in U'_n}2^{-(f(x)+c)+n}\leq \sum_{x\in U'_n}P'(x)\leq 1.\]
Let \(P^{\prime\prime}(x):=2^{-(f(x)+c)+n}\) for \(x\in U'_n\) and \(0\) otherwise. 
Since \(f\) is computable and \(x^\infty\in\cap_n\tilde{U}'_n\), by applying Shannon-Fano-Elias coding to \(P^{\prime\prime}\), we have
\begin{align*}
\exists c_1,c_2 \forall n\exists x\sqsubset x^\infty\ Km(x) &\leq  f(x)-n+2\log n+c_1\\
&\leq -\log P(x)-n+2\log n+c_2,
\end{align*}
where the last inequality follows from (\ref{log-approx}).

Conversely, let 
\begin{align*}
\forall n\ U_n &:=\{x\mid Km(x)<-\log P(x)-n\} \text{ and }\ V_n :=\{x\mid Km(x)<f(x)-n\}.
\end{align*}
From (\ref{log-approx}), we have
\[\exists c\forall n\  V_n\subseteq U_n\subseteq V_{n-c}.\]
Since \(f\) is computable, \(\{V_n\}_{n\in\Nb}\) is uniformly r.e.,   and 
\[\forall n\ \ P(\tilde{V}_n)\leq P(\tilde{U}_n)\leq \sum_{x\in U'_n}2^{-Km(x)-n}\leq 2^{-n},\]
where \(U'_n\) is a prefix-free set such that \(\tilde{U}'_n=\tilde{U}_n\).
Therefore \(\{V_n\}_{n\in\Nb}\) is a test. Since \(\cap_n\tilde{V}_n=\cap_n\tilde{U}_n\), we have the theorem. 
\qed

\begin{center}
Acknowledgement
\end{center}
This work is supported by JSPS KAKENHI Grant number 24540153.


{\small
\begin{thebibliography}{1}

\bibitem{Laurent-etal}
L.~Bienvenu, P.~Gacs, M.~Hoyrup, C.~Rojas, and A.~Shen.
\newblock Algorithmic tests and randomness with respect to a class of measures,
  2011.
\newblock arxiv:1103.1529v2.

\bibitem{kjoshanssen}
Bj{\o}rn~Kjos Hanssen.
\newblock The probability distribution as a computational resource for
  randomness testing.
\newblock {\em Journal of Logic and Analysis}, 2(10):1--13, 2010.

\bibitem{levin73}
L.~A. Levin.
\newblock On the notion of a random sequence.
\newblock {\em Soviet.~Math.~Dokl.}, 14(5):1413--1416, 1973.

\bibitem{LV2008}
M.~Li and P.~Vit{\'a}nyi.
\newblock {\em An introduction to Kolmogorov complexity and Its applications}.
\newblock Springer, New York, third edition, 2008.

\bibitem{schnorr73}
C.~P. Schnorr.
\newblock Process complexity and effective random tests.
\newblock {\em J.~Comp.~Sys.~Sci.}, 7:376--388, 1973.

\bibitem{schnorr77}
C.~P. Schnorr.
\newblock A survey of the theory of random sequences.
\newblock In Butts and Hintikka, editors, {\em Basic problems in Methodology
  and Linguistics}, pages 193--211. Reidel, Dordrecht, 1977.

\bibitem{takahashiIandC2}
H.~Takahashi.
\newblock Algorithmic randomness and monotone complexity on product space.
\newblock {\em Inform.~and Compt.}, 209:183--197, 2011.

\bibitem{takahashi2014}
H.~Takahashi.
\newblock Generalization of van lambalgen's theorem and blind randomness for
  conditional probabilities, 2014.
\newblock arxiv:1310.0709v3.

\bibitem{USS90}
V.~A. Uspenskii, A.~L. Semenov, and A.~Kh. Shen.
\newblock Can an individual sequence of zeros and ones be random?
\newblock {\em Russian Math.~Surveys}, 45(1):121--189, 1990.

\end{thebibliography}
}
\end{document}